\documentclass[12pt]{article}
 \usepackage{amsfonts}
 \usepackage{amssymb,amsmath}
 \usepackage{graphicx} \usepackage{epstopdf}
 \newcommand{\crlb}[1]{\label{#1}\\[2pt]}
 
 \newcommand{\crld}[1]{\label{#1}}
 \newcommand{\eela}[1]{\quad\hbox{\scriptsize{#1}}\label{#1}\end{eqnarray}}
 \newcommand{\eelb}[1]{\label{#1}\end{eqnarray}}
 
 \newcommand{\newsecb}[2]{\section{#1}\label{#2}\setcounter{equation}{0}}

 \newcommand{\nolabels} {\def\eel{\eelb}\def\eeql{\eeqlb}  \def\crl{\crlb} 
 \def\newsecl{\newsecb}\def\bibiteml{\bibitem} \def\citel{\cite}\def\labell{\crld}}
\newcommand{\eeqla}[1]{\quad\hbox{\scriptsize{#1}}\label{#1}\end{aligned}\end{equation}}
\newcommand{\eeqlb}[1]{\label{#1}\end{aligned}\end{equation}}

\newcommand\publishversion  {\nolabels\setlength{\textheight}{8.3in}\setlength
    {\oddsidemargin}{0in} \setlength{\textwidth}{6.3in}\setlength{\topmargin}{-0.2in}}

\def\beq{\begin{equation}\begin{aligned}}		\def\eeq{\end{aligned}\end{equation}}
\def\be{\begin{eqnarray}}  					\def\ee{\end{eqnarray}}		
   \def\bi#1{\begin{itemize}\item[#1]} 	     \def\itm#1{\item[#1]} 	   \def\ei{\end{itemize}} 
   \def\eqn#1{(\ref{#1})}
   	 \def\fn{\footnote}	\def\nm{\nonumber}

\def\m{\mu}	    		        		          
 	 		\def\s{\sigma}     	      	 
	     		          		              	  
    		  		\def\dd{{\rm d}}

          \def\pa{\partial}		\def\ra{\rightarrow}	
 		\def\ket{\rangle}

\def\fract#1#2{{\textstyle\frac{#1}{#2}}}	 	 	
\def\ffract#1#2{\raise .2 em\hbox{$\scriptstyle#1\,$}\kern-.3em/\kern-.2em\lower .15 em \hbox{$\scriptstyle\,#2$}}
\def\half{\fract12}			\def\halff{\ffract12}		
 
\def\ex#1{e^{\textstyle#1}} 			
\def\bpmatrix{\begin{pmatrix}} 			\def\epmatrix{\end{pmatrix}}
\def\bmatrix{\begin{matrix}} 			\def\ematrix{\end{matrix}} 
\def\bcenter{\begin{center}}			\def\ecenter{\end{center}}

\def\lowerheightfig#1#2#3{\(\raise-#1\hbox{\includegraphics[height=#2]{#3}}\)}
\def\lowerwidthfig#1#2#3{\(\raise-#1\hbox{\includegraphics[width=#2]{#3}}\)}

  \def\ds{\displaystyle} 
\def\weglaten#1{}	
 \def\twomat#1#2{\Big(\begin{matrix} #1 \\ #2 \end{matrix}\Big)} 
 \publishversion

\begin{document}

\begin{titlepage}
 \title{Fast Vacuum Fluctuations and the Emergence of Quantum Mechanics}
\author{Gerard 't~Hooft}

\date{\normalsize
Faculty of Science,
Department of Physics\\
Institute for Theoretical Physics\\
Princetonplein 5,
3584 CC Utrecht \\
\underline{The Netherlands} \\[10pt]
e-mail:  g.thooft@uu.nl \\ internet: 
http://www.staff.science.uu.nl/\~{}hooft101}

 \maketitle

\begin{quotation} \noindent {\large\bf Abstract } \\[10pt]

Fast moving classical variables can generate quantum mechanical behavior. We demonstrate how this can happen in a model. The key point is that in classically (ontologically) evolving systems one can still define a conserved quantum energy. For the fast variables, the energy levels are far separated, such that one may assume these variables to stay in their ground state. This forces them to be entangled, so that, consequently, the slow variables are entangled as well. The fast variables could be the vacuum fluctuations caused by unknown super heavy particles. The emerging quantum effects in the light particles are expressed by a Hamiltonian that can have almost any form.

The entire system is ontological, and yet allows one to generate interference effects in computer models. This seemed to lead to an inexplicable paradox, which is now resolved:
exactly what happens in our models if we run a quantum interference experiment in a classical computer is explained. The restriction that very fast variables stay predominantly in their ground state appears to be due to smearing of the physical states in the time direction, preventing their direct detection. Discussions are added of the emergence of quantum mechanics, and the ontology of an EPR/Bell Gedanken experiment. \end{quotation}\end{titlepage}
 
	\newsecl{Formulating our problem} {intro.sec}\setcounter{page}{2}
It seems to be reasonable to demand that the huge successes of the quantum mechanical machinery in describing the statistical features of atoms, molecules, and elementary particles, should be explained in terms of simple ontological descriptions of what is going on in these sub-microscopic objects. Yet there appear to be  problems standing in the way of further exploration of this idea. First, attempts to construct explicit models that explain the origin of Schr\"odinger's equation, invariably ended in unwieldy constructs, either requiring `pilot waves\,\cite{Bohm-1952.ref}' inundating infinitely many virtual universes\,\cite{Everett-1957.ref,DeWitt-1970.ref}, or non-linearities\,\cite{GRW-1986.ref,Khrennikov-2014.ref} that seem to jeopardise the fundamental and perfect beauty of the Copenhagen picture.\fn{For an overview of the history on the  meetings in Copenhagen and elsewhere that resulted in an important consensus, see e.g. A.~Pais\,\cite{Pais-1986.ref}.} It is as if the best thing we can do to-day is to be `agnostic' about the origin of quantum mechanics.
 
 It was felt that the ontological description of quantum mechanics\ requires an extra `axiom', referred to as the collapse of the wave function. Also, the need was felt for a proof that justifies the identification of probabilities with the absolute squares of amplitudes. Why are certainties replaced by probability distributions in quantum mechanics? 

Ignoring all difficulties that an ontological picture of quantum mechanics appears to lead to, some investigators, in particular computer scientists, continued to speculate that the universe may be regarded as a gigantic, classical,  \emph{cellular automaton}\,\cite{Zuse.ref,fredkin.ref,Wolfram-2002.ref,Wolfram-2020.ref}.

The author has repeatedly claimed\,\cite{GtHCA.ref,GtHonto.ref,GtHdetmath.ref}  that models do exist, where both the full quantum machinery applies -- without any deviation or approximation -- and at the same time a totally classical formalism is valid. These models are fundamentally simple, and no `collapse axiom' is needed at all. The probabilities can be naturally identified with the absolute squares of the wave functions, a fact that can be used as input for a mathematical treatment, rather than something that has to be proved or conjectured. These points have recently also been emphasised by Wetterich\,\cite{Wetterich.ref}.

A subclass of classical, linear, discrete cellular automata was discussed further by Elze\,\cite{Elze.ref}. It is not clear to the present author how his earlier approach can accommodate for interactions, but it is conceivable that second-quantisation procedures can be applied from such starting points as well. Elze's later publication\,\cite{Elze2.ref} is more in line with the CA philosophy of the present paper; careful examination of his examples will also exhibit the problem that we now address: it looks as if all Hamiltonians obtained through these lines should  feature sequences of equidistant energy eigenstates, and this incompatibility with the observed quantum systems seems not to go away when interactions are introduced. Exactly here the \emph{fast variables} should come to the rescue, as we shall demonstrate.

New and related problems gave rise to objections that are actually quite legitimate.\cite{TPalmer.ref} One of these was the apparent contradiction with Bell's theorems\,\cite{Bell.ref,Bell1982.ref,Bell1887.ref}. Bell's theorems however required the consideration of statistics, while our models are essentially formulated at a level where we can do away with statistics entirely -- they only make exact statements about states, which are either true or false, but nothing in between. Probabilities in the formalism exclusively originate from the adagium ``probability in = probability out", or, whatever one puts in the initial state will determine what one finds in the final state. Notions such as 'free will'  
\,\cite{CK-2008.ref,Zeilinger-2010.ref} and `superdeterminism'\,\cite{superdet-2011.ref,Hossenfelder-2014.ref} do not mean anything here.  

Bell's definitions of `causality' or `statistical independence' do not apply. These points are eloquently worded by Vervoort\,\cite{Vervoort-2013.ref, Vervoort-2019.ref}. The most surprising thing one concludes from these papers is that elaborated discussions seem to be needed to defend the very idea that ``every event must be caused by earlier events, a causal relationship that goes back all the way to the Big Bang." Indeed this was denied by Bell (and many of his followers up to this day), on the basis that `free will' should exist.  On social media one reads statements such as: `if one would have to resort to believing in super-determinism, this would mean the end of science'. In our paper we take the view point that suspecting correlations between the (not directly observable) wave functions of photons in the past, and the settings chosen by Alice and Bob in the future, may not implicate the end of science. It just requires a more detailed discussion of the interpretation of quantum mechanics: how `ontological' is a wave function? How ontological is a probability, if it may be correlated with what happens in the future? May a wave function in the past have a butterfly effect on the future?

Our point will be that models are only deterministic if the values of all dynamical variables are exactly accounted for. One is certainly allowed to consider statistical arguments, but they have to be cast in a quantum mechanical language. This we shall explain in the discussion sections at the end of the paper. The surprising feature of our findings is that classical probabilities can be treated using quantum mechanics as a tool, not as a theory, in a mathematically legitimate fashion. Quantum mechanics, without the slightest modification from the Copenhagen doctrine, may be regarded as an `emergent' theory.

More to the point is an objection that the author has been struggling with more recently, which is that we should be able to construct precise models that reflect how phenomena typical for quantum mechanics, but not for ordinary classical systems, can come about. What is it that `really happens' in a `Gedanken experiment' where Bell's theorems, or the well-known CHSH inequalities\,\cite{CHSH.ref} are violated? What happens when a quantum interference experiment is performed in a (classical) computer?

We added an answer to this question in Appendix\,\ref{EPRBell.sec}.
Constructing explicit models is technically hard, but possible in principle. The principle was explained in Ref.~\cite{GtHCA.ref}: all one has to do is re-formulate the quantum system one wishes to investigate in such a way that wave functions exclusively take the values 1 or 0. In that case, we are no longer dealing with probabilities but only with certainties: \(\psi=1\) means `yes, this is the truth', and \(\psi=0\) means `no this isn't it'. Realising a transformation  that turns a quantum system in such an unrecognisable form, turns out to require many unitary rotations in Hilbert space.  So be it, the truth is somewhat hidden from us because we have been taught emphatically how to use probabilistic descriptions of everything we attempt to describe, whenever quantum mechanics has been adopted.

As soon as we found wave functions taking the values 1 and 0 only, we also find that the evolution equations transport realities (\emph{beables}) into other realities. This means that these evolution operators turn into elements of a mundane permutation group. And this is classical physics. Yet, since elements of a permutation group can certainly be represented in terms of unitary operators, this \emph{is} quantum mechanics at the same time.

Now the question is how such models can be found, or, which transformations transform the presently familiar quantum models into the much more restricted wave functions just described. The new descriptions hardly look like quantum mechanics as it is usually known, yet they accurately obey the quantum mechanical rules where probabilities are replaced by certainties, as one might think scientific descriptions should be, at their most basic level. 

To realise this, we start with a completely classical system. Indeed, its evolution law formally consists of elements of the permutation group. We re-write it in quantum mechanical \emph{language} using exclusively wave functions of the form \(\psi=1\) and \(\psi=0\).  In order to re-write the model in such a way that its Schr\"odinger equation reproduces a Hamiltonian of a pre-assigned form, we first assume the existence of very high frequency oscillations. These give rise to energy levels way beyond the regime of the Standard Model. When now these fast oscillating quantities are handled as classical variables that have to be smeared over short, finite periods of time, one obtains the familiar quantum mechanical expressions, allowing for operators showing the structures that we are used to. We end up not just with a system that mimics quantum mechanics, but with the real thing.

Along these lines, one can obtain any quantum Hamiltonian, as will be shown, except that in our models the spacings of the energy levels are not completely free; they are constrained in terms of rational numbers. It is not hard to argue that, in a realistic model of the microcosmos, the distinction between rational and real numbers will be unnoticeable in practice, after having taken a large volume limit and/or  a high complexity limit.

In short, what our procedure amounts to, is that we transformed our classical system back to any of our familiar quantum models, by limiting our observations to states that are very slightly smeared in time. We emphasise that this shows that the distinction usually made between quantum systems and classical systems is, to a large extent, illusionary. One can also say that quantum mechanics is what one gets if probabilistic considerations are combined with allowing ultrafast evolving variables, in a setting that starts out being  basically completely classical.

We do note that not all classical systems allow for a treatment that turns them into quantum systems without affecting the underlying physics; it may only happen if, at its most basic, small-distance, level, our system is discrete.\fn{And this implies that the words `\emph{quantum mechanics}' are actually aptly chosen: variables that start out being chosen to be continuous, are actually discrete (quantised). Had the theory been called `uncertainty dynamics' or something like that, we would have been forced to rename it.} This discreteness implies that our systems have far fewer distinct orthonormal states than conventional classical systems such as a Van der Waals gas or a globular star cluster,

We do not here perform all technical calculations all the way.  We do show how realistic models can be constructed, and we suspect that our results might one day be utilised to help us guess what the sub-atomic world will look like beyond the domain explored by the latest particle accelerators such as  LHC, and how to connect our particle models with what is known about cosmology, a science that partly covers uncharted territory of high energy physics.

It will be important to note that our considerations do not require non-locality, but we do have problems explaining local gauge invariance, special and general relativity, and other continuous symmetries. This does not mean that these symmetries would disqualify our approach, but it means simply that we have not yet reached the stage that we can understand the origin of the physically observed symmetries in our classical models. These belong to the many important questions that have to be left for the future.

In the present paper we show how to make models that separate fast variables from slow ones, and how energy conservation (in particular for the fast variables), can be exploited to explain why our world seems to be perfectly quantum mechanical, while it appears to be difficult to identify the original ontological variables. A more rudimentary description of our models was also given in ref.~\cite{GtHdetmath.ref}.

The classical system we start with is absolutely classical, hence it does away with collapse axioms, measurement problems, pilot waves and many worlds.\fn{One could argue that our fast fluctuating variables resemble pilot waves in some mathematical sense, and the large number of states forcing them to behave as white noise may generate some resemblance with a many world picture, so a conservative interpretation of our findings could be that we now found a mathematical procedure that places these features in a physically much more acceptable setting.} All we need is fast variables such as the vacuum fluctuations due to heavy particles, many of which being unknown today.

The fast variables are moving completely classically, but too fast to be followed by the `observers' in our model. The dynamical equations of motion  are assumed to be very simple,\fn{Since the fast variables tend to form a stochastic background, our approach has similarities with Nelson's stochastic theories of quantum mechanics\,\cite{Nelson.ref}. However, we claim to have a more natural explanation than Nelson, as to the origin of Schr\"odinger's equation. We insist that the behavior of these `stochastic' variables is also deterministic and time-reversible. Our preference for \emph{fast} variables rather than stochastic ones, arises from insisting that these variables should not generate unphysical quantum states, otherwise, it is difficult to see that effectively quantum operators emerge, in particular quantum interference, see Appendix  \ref{interfer.app}. }  and they are explained in section~\ref{fast.sec}. The slow variables will be assumed to perform flip-flop movements, but before describing them, we digress to explain the ontological significance of the use of complex numbers in quantum mechanics.  In Section~\ref{bits.sec}, we explain  why, in conventional quantum mechanics, our evolution operators are usually unitary rather than orthogonal. 

The slow variables evolve classically through switches, which are dictated by the fast variables. Their equations are described in section~\ref{slow.sec}. Then comes the most essential part of this paper (Section~\ref{quant.sec}): just because the fast variables are in, or close to, their lowest energy eigenstates, the Hamiltonian for the slow variables turns into a quantum mechanical one, hiding its classical origin. We shall elaborate what it means to have the fast variables in energy eigenstates, while they are still classical: they merely have a statistically smooth distribution, which today's physical detectors are unable to disentangle. They subsequently send the slow variables into entangled states as well. These variables also are classical, but it is our quantum mechanical machinery that produces statistical distributions of the slow variables such a way that they seem to be quantum-entangled.

The result is discussed in section~\ref{disc.sec}. Earlier models raised the question how one can account for quantum interference effects. How do the ontological variables turn into conventional quantum operators and observables that come in wave functions that can be superimposed? How can interfering waves of particles be simulated in a classical computer? This seemed to give rise to contradictions. A flash of insight produces the correct answer to such questions. Only one set-up is treated in this section: the two-slit interference experiment. Other, related questions must be answered in a similar fashion. In this renewed version of this paper, the discussion of interference is referred to in appendix (\ref{interfer.app}), and also some other issues are handled in Appendices, being the discussion of symmetries, which may be helpful for understanding how the Standard Model can be linked to cellular automata (Appendix~\ref{entsym.app}), and some conceivable implications for quantum computing (Appendix~\ref{qucomp.app}).

\newsecl{The fast variables}{fast.sec}

We now discuss the generic structure of a class of fundamental models  in  more detail.  If one would treat all variables, fast and slow, at the same level, one would be able to derive a Hamiltonian that at first sight indeed would look like a fully quantum mechanical one. However, there is one fundamental problem: its energy levels then have conspicuous patterns that we do not see in the real world: the energy levels should form perfect sequences that would feature  exactly regular  spacings everywhere. This is described in Ref.~\cite{GtHCA.ref}. It is also explained there that this still leaves enough freedom to construct physically interesting models. This is formally true, but in practice it is an impediment. We are offering a more practical way out: it is 
 suggested that strictly linear sequences of energy levels will be characteristic primarily for the invisible fast fluctuating variables. Their `exactly equal separations' are invisible to us because the actual energy spacings are much greater than a TeV, or whatever the energy is, up to which the Standard Model could be checked. 

This implies that, yes, we can ascribe the quantum features of our world to classical evolution equations, but only if these classical laws act at ultra-short time scales.

And what about the slow variables? Do they contain ladder-like sequences of energy levels? It will turn out that they consist primarily of binary variables, pairs of states that flipflop into one another. This gives them just pairs of energy levels, so no conspicuous sequences will emerge. The fact that the slow variables nevertheless evolve in a non-trivial way will be seen to be due to their interactions with the fast variables.

The qualitative description above, of our models to be constructed, helps us now in formulating the basic equations. The fast fluctuating variables are taken to live on periodic spaces, typically forming a multi-dimensional torus with very tiny physical length scales. One could regard these as `extra dimensions'\,\fn{We put `extra dimensions'  between quotation marks as there is no need for extending the Lorentz group to these extra coordinates.}. A few of such extra dimensions suffice:  our model assumes one fast periodic variable for each possible quantum state of the slow  variables, so one can think of a (classically evolving) field living in space-time with just a few extra dimensions added.  The slow variables just live in ordinary space-time (the `3-brane').

We take the number of fast variables\fn{In more sophisticated models in the future, the number of really fast variables needed might be a lot smaller.}, and the dimensionality of the Hilbert space containing the slow variables, both to be  \(N\). This number therefore counts all normally observable states the system can be in. In models for the entire world, \(N\) of course is extremely large, but the principle holds all the same. The number of variables inside a given small volume of space (think of a volume of Planckian dimensions), may be kept small. Thus \(N\) will be the volume of the universe -- or the part of the universe described by our model -- in Planck units. 

Here are the equations for the fast variables of our model.
The classical evolution law is:
\be x_i(t+1)=x_i(t)+1\mod L_i\ , \eel{classevolv.eq} for all values of the index \(i=1,\,\dots,\,N\).
It can be addressed by casting its laws in the form of an ordinary Hamiltonian, such that the evolution agrees with a Schr\"odinger equation:
\be H_0=\sum_ip_i\ ,\qquad p_i=-i \frac{\pa}{\pa x_i} = \frac{2\pi n_i}{L_i}\ , \qquad n_i=0,1,\cdots,\,L_i-1\ .\eel{fastH.eq}
One easily checks that, at integer values of the time \(t\), this Hamiltonian moves all variables \(x_i\) across integer points of a lattice on the torus. The velocities are all the same. It will be important to have this lattice here (rather than a continuum), otherwise difficulties may arise in defining exactly what the classical states of the system are.\fn{In a more precise formalism, one thinks of having exactly one strictly continuous variable, running around in circles, while all others are discrete, Such a structure allows us to have a strictly discrete Hilbert space at any given time,  while nevertheless time can be defined to be a continuous variable, enabling us to formulate the quantum evolution in terms of a Schr\"odinger equation        
equipped with a Hamiltonian. This one special periodic variable is then linked to the binary variable we call \emph{c-bit}, in Section~\ref{bits.sec}.}
Our later considerations will require that the (large) numbers \(L_i\) will, to a large extent, be relative primes (section~\ref{slow.sec}).

 The spacings between the energy levels form a simple grid. Actually, our definition of the Hamiltonian was assuming continuous variables \(x_i\)\,, with respect to which one can differentiate.  It is easy however to find the Hamiltonian when we restrict ourselves to the integer lattice points only, by deriving the eigenstates of the evolution law \eqn{classevolv.eq}, which are found to agree with the Hamiltonian \eqn{fastH.eq}. On the discrete lattice, however, one finds one further restriction: since, in any direction \(i\), there are only \(L_i\) states -- a finite, integer number --, the momentum quantum numbers \(n_i\) can all be restricted to \emph{non-negative} integers as given in eq.~\eqn{fastH.eq}.

 The ground state is then found to be a wave function that is constant over the entire torus, so that its energy is zero. All excited states are way beyond a TeV simply because the torus is so small. Thus, if we would live in such a model, we would have no knowledge today of these extra states, just as we have no knowledge of new particles with masses far above a few TeV; we now assume them to exist.

To excite any of these higher energy states,  more energy would be required than the quantum energy we can generate in any Earth bound laboratory, and therefore it sounds reasonable that we ignore them all. This is how one usually deals with `unknown high mass particles' in  `Grand Unified Theories'.  Here it is the key ingredient of the mechanism that we propose. There may well be reasons to doubt whether our mechanism will work as expected, but a comparable case is for instance the decay of a proton (a slow process), which can be ascribed to an exchange of a very high mass vector boson (a high energy excited state).

We now consider our two domains of physical states. Besides the states of the fast variables described above, we have \(N\) states that we can indeed observe. At this point, the Hamiltonian \eqn{fastH.eq} does not depend on these states, and consequently all energy levels of the fast variables, discussed above, are degenerate with multiplicity \(N\). As yet, these low energy states do not evolve.  It is these \(N\) states that we consider to be observable, in the quantum mechanical sense; they are taken to be ontological also, \emph{but they are only ontologically observable for observers who also monitor the fast variables,} otherwise, they will be ill-defined, and this will be shown to be how they get to become quantum observables.

Our next step will be the description of the domain of slowly varying states, where we claim quantum mechanics will be spontaneously generated. But first, we need to describe how ontological binary data can be made to flip in various ways. 

The more experienced reader might want to skip the next section, proceeding immediately to Section~\ref{slow.sec}.

\newsecl{Qubits, c-bits, and the sign of the wave function}{bits.sec}

We saw that for  a system that evolves classically, as in eq.~\eqn{classevolv.eq}, one may introduce a wave function that obeys the Schr\"odinger equation \(\frac{\dd}{\dd t}|\psi\ket_t=-iH_0|\psi\ket_t\).
But actually, this Schr\"odinger equation may seem like overkill: the system shifts position as in eq.~\eqn{classevolv.eq}, but wave functions do not really mix, they are just transported as a whole. This means that, at this point, Born's rule\,\cite{Born-1926.ref} is trivially valid, when we merely define \(|\psi|^2\) to represent probabilities. Whatever phase we add to these wave functions, is simply transported without anything happening to it.

Yet phase will mean something quite special in ordinary quantum mechanics. We shall attribute all of this to the mathematical procedures that will be applied, but let us first define the wave functions more precisely. Consider the phase. It is due to our addiction to complex numbers; however, the phase will stand not only for addiction, but also something real: \emph{a complex number is a pair of real numbers.} 

This implies that, when we have a complex wave \(\psi=\alpha +i\beta\), we can interpret this as a system that can be in two classical states. Therefore, we observe that conventional quantum states contain one special kind of binary object, to be called a \emph{c-bit} (`c' standing for `classical' or `complex-number'), which is a state that can be in position \(\Re\) or in position \(\Im\):
\be|\psi\ket = (\alpha +i\beta)\,|.\ket=\alpha |\Re\ket+\beta|\Im\ket\ ,\qquad \alpha \ \hbox{ and }\ \beta\ \hbox{ real}\ . \ee
Taking this c-bit into account, our wave function is now a real number.  The probabilities of having state \(\Re\) or state \(\Im\) are \(\alpha ^2\) and \(\beta^2\), respectively.

This gives us an almost unique definition of the wave function in terms of probabilities: 
\emph{the wave function is uniquely defined to be plus-or-minus the square root of the probability,}
but what does its sign stand for?

As far as we know, there is only one c-bit in the universe, so it really serves a purely mathematical purpose. Having this c-bit is necessary however, since it indicates something that should be observable in the technical sense (even if the world fluctuates strongly between its two states \(\Re\) and \(\Im\))\,\fn{Only in a state with total energy zero, the c-bit is conserved in time, as one can easily check.}. The number \(i\) is actually an operator here, exchanging the states \(|\Re\ket\) and \(|\Im\ket\):
	\be i\twomat{\alpha \,}{\beta}=\twomat {-\beta\,}{\alpha}=\twomat{0&-1}{1&0}\twomat{\alpha\,}{\beta}\ . \eel{i.eq}

This sign is crucial if we wish to replace discrete time steps by continuous time steps. 
If time is discrete, then after every step the system flip-flops from \(|\Re\ket\) to \(|\Im\ket\) and back. In the continuum description, we wish to describe this as a rotation over \(90^{\circ}\). We cannot keep plus signs everywhere. Instead of making finite, discrete, replacements, we need to make infinitesimal additions to describe the evolution, but then we need subtractions as well. If we use the correct ontological basis, the additions add the desired state and the subtractions remove the unwanted states.\fn{If all probabilities are limited to certainties, then the only allowed wave functions are \(+1,\,0,\) and \(-1\), so subtraction means replacement by 0.} Now, if we switch the sign of the imaginary part of the wave function, it will continue to evolve, but backwards in time. Therefore, the importance of the relative sign of \(|\Re\ket\) and \(|\Im\ket\) is that it indicates whether our evolution law will transport it forward or backwards in time. Switching the sign of \emph{both} the real part and the imaginary part has no direct physical aspect for the ontological theory.\fn{There is an other way to interpret the c-bits \(\Re\) and \(\Im\), namely as the sine and cosine of an angle. As we indicated in a previous footnote, to describe discreteness in quantum physics, one can replace all variables by discrete variables, \emph{except for the one variable the monitors the progress of time}. Even in discrete systems, it is useful to keep one continuous variable for time, running in a circle, rather than a discrete clock. That's the c-bit. It helps us for instance to define the Hamiltonian, without modifying the physical essentials.}

Consider now any other binary observable.
Suppose it makes a switch, taking just a short time to do this, typically \(\delta t\approx 1\) in the time units used in eq.~\eqn{classevolv.eq}. Consider a time dependent Hamiltonian. Use
	\be \ex{\pm \pi i/2}=\pm i\ ,\qquad \ex{\pm \pi i}=-1\ , \ee
to prove that for any operator \(\s\) that has all eigenvalues equal to \(\pm 1\), one has
\be \ex{\fract{\pi i}2 \s}=i\s\ ,\qquad\ex{ \pi i \s}=-1\ . \ee
	We find that the Hamiltonian
		\be H(t)=\half \pi \s\delta (t-t_1)\ ,\eel{sigmaHam.eq}
yields the evolution operator \(U_{t_1}\) from \(t<t_1\) to \(t>t_1\) equal to
		\be U_{t_1}=\ex{-i\int_{t<t_1}^{t>t_1} H(t)\dd t}\ =\ -\,i \s\ . \ee
For the c-bit, this tells us that a switch from \(|\Re\ket\) to \(|\Im\ket\) is generated by the (time dependent) Hamiltonian\,\fn{The continuum notation is used here. For the discrete case, the Dirac delta function may be replaced by a Kronecker delta, with some due care.}
\be H=\half\pi\twomat{0&i}{-i&0}\delta(t)\,. \eel{c-bitHam.eq}

Any other binary operator such as a spinor, consists of two states, each to be covered by a complex wave function. Thus there are 4 real numbers. This is traditionally called a qubit. As its wave function is complex, our system of operators acts on two binary states, that is, one c-bit and one other binary object describing spin or anything else. Let us assume the spin in the 3-direction to be \(\pm\half\). The states are now
\be \{1,\,2,\,3\,,4\} \ = \ \big\{\, |\Re,\,+\half\ket,\quad |\Im,\,+\half\ket,\quad |\Re,\,-\half\ket,\quad |\Im,\,-\half\ket\,\big\}\ .\ee
These are the four ontological states. Ontological operations on these states consist of the \(4!=24\) possible permutations, with in addition some sign switch operators. There is only one relevant sign switch, telling us whether the evolution will proceed forwards or backwards in time.\fn{The relevant sign switch is \emph{complex conjugation}, that is, giving \(\Im\) but not \(\Re\) a minus sign.} Thus, there are \(2\cdot 4!=48\) distinct ontological operations.

These we can consider as built up from interchange operators. The most important ones\fn{other members of the class of switches are obtained by multiplication with \(\halff(1\pm \s_a)\), which means that the states with \(\s_a=+1\) or \(-1\) are singled out.} for us are generated by the Pauli matrices, \(\s_1,\,\s_2\,,\) and \(\s_3\). If we substitute these in our Hamiltonian
\eqn{sigmaHam.eq}, we find the ontological switches realised by \(\s_a\) and \(i\) to be
\be		i\s_1=\twomat{0&1}{-1& 0}_c\,\twomat{\,0&1}{\,1&0}_s ,&\ & i\s_2=\twomat{0&1}{-1&0}_s\ ,
		\labell{sigma12i.eq}\\
		i\s_3=\twomat{0&1}{-1&0}_c\,\twomat{\,1&0}{\,0&-1}_s\ ,& & i= \twomat{\,0&-1}{\,1&0}_c\ , \eel{sigma3i.eq}
where the subscript \(c\) stands for the c-bit and \(s\) for the spinor.

We see that \(\s_1\) simultaneously switches the spinor and the c-bit, \(\s_2\) switches only the spinor, and \(\s_3\) only switches the c-bit. The product of two Pauli matrices does the same as the third one. \(i\) does nothing physical, it redefines the floor of the Hamiltonian.

This exercise was made to explain why all three Pauli matrices generate ontologically distinguishable switches, while even the spin-independent Hamiltonian (second part eq.~\ref{sigma3i.eq}) corresponds to repeated switches from \(\Re\) to \(\Im\) and back, as one could have expected from the solutions for the spin independent Hamiltonian.

Ontological models should only consist of operators of the form \eqn{sigmaHam.eq}, where only one out of a small set  of transitions is allowed at any time \(t_1\), and the constant \(\half\pi\) is fixed, otherwise we produce superpositions, which was not allowed in our classical models. Note that the switches in eqs.~\eqn{sigma12i.eq} and \eqn{sigma3i.eq} are all antisymmetric matrices.\fn{Side remark: we could rephrase the theory of quantum mechanics for a theory with only real wave functions by replacing the Hamiltonian by a real, antisymmetric matrix. This cancels out the \(i\) in eq.~\eqn{c-bitHam.eq}, so that real functions evolve into real functions only\,\cite{Finkelstein.ref}. It makes little difference in physics, actually simplifies things somewhat, but I found the formalism rather unfamiliar, requiring further explanations on the way, so we shall not hang on to it.}

\newsecl{The slowly evolving states}{slow.sec}

In section~\ref{fast.sec}, we ended with having \(N\) non evolving states in addition to our fast variables. It is these \(N\) states that will be supposed to describe our world, in terms of slow variables. If we consider fast variables and slow variables together, their joint evolution will be demanded to be totally classical -- quantum mechanics will emerge if we eliminate the fast variables, as will be shown below.  The quantum mechanical system that we shall arrive at can have an almost arbitrary pre-set effective quantum Hamiltonian.

The Hamiltonian we have up to now, \(H_0\), eq.~\eqn{fastH.eq}, does not directly affect the slow states \(\psi_i=|i\ket\) at all, where \(i\) runs from 1 to \(N\). Now, we add an interaction that changes this. It is a switch operation, as defined in section~\ref{bits.sec}. Consider two states, \(|i\ket\) and \(|j\ket\), with \(i<j\), and impose classically that these two states switch one into the other, at the moment \(t_1\), but it is arranged to occur only when both the (discrete) fast variables \(x_i\) and \(x_j\) happen to be at given positions \(x_{i,1}\) and  \(x_{j,1}\):
 	\be H_{1,ij}=\pm\half\pi\s_a(i,j)\,\delta(_{\ds x_i,x_{i,1}}\,\delta_{\ds x_j,x_{j,1}}\ , \eel{Ham1.eq}
The deltas are Kronecker deltas. We may replace one of them by a Dirac delta, provided that the corresponding variable \(x\) is then treated as a continuous variable, but we cannot do this for both \(x_i\) and \(x_j\) simultaneously without running into conflicts. For sake of symmetry, we prefer to keep both variables discrete. This Hamiltonian then acts through one unit of time, and one may verify that this requires the unmodified factor \(\half\pi\) in front.\fn{The sign in front of eq.~\eqn{Ham1.eq} has no ontological significance by itself, but it may become important in combination with other switch interactions introduced later in this section. The order in which switches take place then becomes important.}

In eq.~\eqn{Ham1.eq}, \(\s_a(i,j)\) is defined to be the Pauli matrix \(\s_a\), with \(a=1,2,\) or \(3\) acting on the two state world \((|i\ket,\ |j\ket)\)\,. The operator \eqn{Ham1.eq} is the operator \eqn{sigmaHam.eq} in that space. It generates the evolution operators \eqn{sigma12i.eq} and \eqn{sigma3i.eq} in this two state world (plus the c-bit), if the subscript \(s\) is now taken to represent the two states \(|i\ket\) and \(|j\ket\).
Thus, from now on, a binary flipflop takes place whenever \(x_i\) and \(x_j\) simultaneously arrive at the pre-designed values \(x_{i,1}\) and \(x_{j,1}\). So the combination \(H=H_0+H_{1,ij}\) may be used to describe an ontological evolution law.

We may repeat this procedure to add more interaction terms. The evolution stays unitary provided that we use different points \(x_1\) for every term. Classically, the system now hops from one state to another, in response to the classical, fast variables. Nevertheless, this evolution is slow, because the coincidences where both \(x_i\) and \(x_j\) are required to arrive simultaneously at given positions, do not happen often. It typically takes time \(L_i L_j\) for this to happen. Thus, our fast time scale is the largest value of the \(L_i\), the slow time scale is the average product, \(L_i L_j\).

\newsecl{Why is this quantum mechanical?}{quant.sec}
The freedom we have is to choose the pairs \((i,j)\) for which this flipflop may occur, and to choose at which points in the space of the fast variables they occur. The latter choice seems to be of secondary importance since the classical evolution goes very quickly for the fast variables. What matters most is the values of the pairs \((i,j)\) and how frequently switches between given pairs occur.

Now follows an important question: how does the solution of these evolution equations behave? The behaviour is classical. If one tries to solve the equations exactly, one finds that the system stays classical. The energy levels form equally spaced sequences, and consequently, the entire theory seems to be a failure. Imagine however, applying this doctrine to any remotely realistic quantum field theory. Even classically, it will be far too complex to follow with infinite precision what goes on. What we are interested in is only the slow variables \(|i\ket\), and how their behaviour can be described while ignoring the fast variables.

To find good approximated solutions, we now must assume that the fast time scale and the slow time scale are well-separated\fn{In principle, this condition can be relaxed. All that matters is that fast variables are smeared due to limits in our time resolution.}, even if the pairs \((i,j)\) featured in \(H_1\) are fairly numerous. They must still be much fewer than the large numbers \(L_i\) permit. Then, we may employ  the large \(L\) expansion.

At first sight, the evolution of the fast variables is not affected by \(H_1\). Classically, our system is non-Newtonian: there is no reaction of the variables \(x_i(t)\) to what the slow states \(|i\ket\) do, while the slow states \(|i\ket\) are dictated by the fast variables. The quantum Hamiltonian, \(H_0+\sum_{i,j}H_{1,ij}\), is nevertheless fully acceptable, as it is hermitian and finite.

Our approximation will be that the fast variables are in their energy ground state. In spite of the absence of a classical back reaction, there will be a quantum back reaction in the higher energy states, so the classical fast variables will not stay in their ground state. This situation is quite familiar in standard quantum mechanics, and there is no reason to treat the present situation in any way differently from what we usually do: apply perturbation expansions.

This means that all energy eigenstates are shifted due to the perturbation, but the lowest states will still be most frequently used, when energy conservation forbids the higher energy states. The \(N\) slow (classical) states, which started out completely degenerate, now will shift to form some pattern, but one may expect them to stay well separated from the excited energy values for the fast variables. Remember, our model obeys the same type of Schr\"odinger equation as is usual, so there is no need for alarm.

When we calculate the lowest order effect in perturbation theory, we find that, in the Hamiltonian \(H_1\), eq.~\eqn{Ham1.eq}, the \(x\)-dependence for the fast variable must be replaced by its expectation value. Since there are \(L_i L_j\) sites for the fast variable, one therefore must replace the delta functions by their averages, as follows,
	\be \delta_{\ds x_i-x_{i,1}}\,\delta_{\ds x_j-x_{j,1}} &\ra& 1/(L_i L_j)\ ,\nm\\
	H_{\mathrm{slow}}\ \ra\ \sum_{a=1,2,3;\,i<j} H_{1,ij}&=& \sum_{a;\,i<j}(\pm \half\pi)\s_a(i,j)/L_iL_j\ . \eel{effHam1.eq}

The main result reported in this paper is that by adding many interactions of the form  \eqn{Ham1.eq}, the slow variables end up by being described by a fully quantum mechanical Hamiltonian \(H_{\mathrm{slow}}\) that is a sum of  the form \eqn{effHam1.eq}. All switches involving the sites \(i\) and \(j\), together generate the matrix element \(H_{1,ij}\) of \(H_{\mathrm{slow}}\).  It is easy to check that we can reach all possible entries in the Hermitian matrix \(H\).  The diagonal parts of \(H_{\mathrm{slow}}\) are taken care of by the matrices \(\s_3\) and \(i\), the off-diagonal ones by \(\s_1\) and \(\s_2\). The fixed coefficients \(\pm\half\pi\) are now replaced by any arbitrary coefficients, and this has the effect of changing our slow variables from classical to quantum variables. We can mimic almost any quantum Hamiltonian this way, although in the model the strengths of all on- and off diagonal terms are controlled by ratios that have to be rational numbers (they all take the form \(\pi R/(2L_iL_j)\)), where \(R,\ L_i\) and \(L_j\) are integers).

One may also try to evaluate \emph{exactly} what happens: at any of the transition points \(x_{i1},x_{j1}\), the wave functions typically rotate by \(90^\circ\) in some channels. This has no  local effect, it only manifests itself when the boundary conditions on the entire torus are considered, which is exactly equal to the perturbative effect, except when we mix non-commuting contributions; their effects will depend on the order, that is, the exact locations of the transition points. Indeed one can verify that, if all terms of the switches commute, one sees that the exact quantum mechanical effect is obtained:  certainty about the switch is only seen after a full period of the fast variables. In the more interesting case that many non-commuting terms are added in the effective Hamiltonian, the exact solutions are more complicated but still fully quantum mechanical, and accurately reproduced by the perturbative expression \eqn{effHam1.eq}.

The reason why we fail to see longer sequences of energy eigenvalues with much smaller spacings (which is what an exact analysis may seem to give), is that energy is only defined \emph{modulo} the length of the inverse time steps, which results in energy levels of the unperturbed case that are much farther apart then if we considered single rings of length \(L_iL_j\) instead of a torus with radii \(L_i\) 
and \(L_j\). The \(90^\circ\) rotations are still fine in the torus, but technically more difficult to follow.

An interesting question concerns the signs of the effective Hamiltonian terms; there seems to be some freedom in choosing them. Closer inspection suggests that these signs are actually fixed when non-commuting elements are taken into account, but the real mathematical puzzle has not yet completely been resolved.

\newsecl{Discussion}{disc.sec}
We admit that not yet all questions have been answered, but there is a thing that we are quite certain about: in spite of the fact that our theory is entirely ontological, it is also controlled by a 
Schr\"odinger equation, and with the Hamiltonians \eqn{fastH.eq} and \eqn{sigmaHam.eq} all inserted,  this Schr\"odinger equation is exactly valid; it is genuinely quantum mechanical. 

The difficulties many readers appear to have in reading our recent papers, and probably also this one, are due to the fact that, in all conventional discussions, quantum mechanics is dealt with as a \emph{theory}. The present  author however, has learned to understand that quantum mechanics is nothing but a superior way to handle the mathematics of discrete dynamical systems. The equations that are assumed to describe the dynamical evolution laws are entirely classical -- the words Hilbert space and Schr\"odinger equations are not used when these original dynamical laws are defined.

An accurate account of this situation is: we are dealing with a system of states that evolve as dictated by a unitary operator  that happens to coincide with an element of the permutation group. For this reason, the theory is classical: all its allowed states dance to the tune of the permutation group. But the unitary evolution operator can also be written as the exponent of a Hamiltonian operator, and hence it is quantum mechanical as well.

Subsequently, we may decide to make an arbitrary unitary transformation to any other basis of Hilbert space .We may interpret the evolution operator exactly as dictated by Copenhagen. In general, the transformed evolution operator seems to handle probabilities and uncertainties in the usual quantum mechanical way, but now we also observe that  both the initial state and the final state are superpositions of the ontological basis elements. If we interpret the absolute squares of the coefficient amplitudes as probabilities\,\cite{GtHCA.ref}, we get the complete story as given by Copenhagen. This is real quantum mechanics.

We use statistics and probabilities as exact approaches, without allowing for uncertainties in the equations of motion. The only uncertainties allowed are generated by our ignorance concerning the initial states. Given a system of planets, or atoms or radiating fields, one usually has to deal with some randomisation in the initial states. In this paper we just limit the uncertainties entirely to the initial states of the fast variables.

Because of this -- no surprise -- also the final states will form statistical distributions. The relation between initial state and final state, on the other hand, must be determined by the theory with infinite precision, since exact dynamical equations were given. 

What we call quantum mechanics, therefore, is nothing but a mathematical tool. Use is made of wave functions, but we can ordain that also these wave functions describe certainties, not vague distributions, and, in our quantum mechanical language,  this is realised by restricting ourselves to wave functions taking only the values 1 and 0, while also the evolution law does not move them away from the 1s and 0s, to that one finds the evolution operator to consist exclusively of permutations of states -- note that, these permutations do not yet generate minus signs --. But permutation operators can perfectly well be written as unitary operators, and we are allowed to apply completely general unitary transformations to whatever basis of Hilbert space elements we wish to use. If the Born rule is \emph{defined} to generate probabilities out of amplitudes squared, this gives us a superior mathematical scheme for handling probability distributions, \emph{without any modification of the (deterministic) physical laws.}

Phrasing our procedures along such lines, we are left with one more difficult question: which classical system or systems can actually be used to approach the world we see, and in particular: the Standard Model of the elementary particles? How can we match the underlying classical system with all observations that have been made concerning these particles?

We now answer the question by stating that the classical model may not only describe the states we see, but also some hidden variables, which we consider to be \emph{fast, oscillating variables}. These variables could be the (classical) fields of, as yet unknown, super heavy particles.

We decide to average over the initial states of the fast variables in our approach. This is tantamount to postulating the fast variables to be in their lowest energy eigenmodes. Subsequently, we use energy conservation to observe that the excited energy modes will be strongly suppressed at all stages of the evolving system, just because the energy levels of the fast modes are too far separated; the energy to excite them is not available.

Now the energy eigenstates are quantum mixtures of the original pure ontological states, and this is why we soon enough encounter quantum superpositions in our slow variables as well. The entire quantum mechanical machinery applies. It is not ``assumed to apply", it is found to be applicable. Thus we actually derived that quantum mechanics is inevitable when fast moving, discrete but deterministic, variables cannot be followed closely enough to keep these classical states all the way. It is merely an act of efficiency to allow the fast states to get mixed. Only the slow variables are now seen to evolve into superpositions of themselves, as we showed.

Whatever the higher order corrections are that will ensue from our model Hamiltonian, they will merely be small corrections that do not jeopardise the quantum nature of the system.

The real reason why we have evaded the usual no-go barriers is something more subtle: we imposed that the higher energy states must be forbidden by the law of energy conservation. It is the \emph{total} energy that is constrained to be small. ``Small" here means small in he units of the fast variables; the constraint is insignificant in the units of the slow variables, so that, for the slow variables, the first \(N\)  energy eigen states contribute.

Via thermodynamics, this also generates  statistical dominance of these (relatively) low energy states locally. Thus we effectively added the \emph{total energy} to our set of ontological observables. This is a new twist to our story. Energy does not commute with our other ontological observables. We know from  practical experiences with quantum mechanics however that energy is observable, but only if other ontic observables are smeared over time. Now since we talk of very \emph{large} energies in the fast variable, we will be dealing with time smearing over \emph{tiny}, hence  in practice insignificant, amounts of time.

In terms of our classical observables, the quantum  notion of  energy  does not exist. When we say that the fast variables are in their energy ground state, what this really means is that the probability distribution of these fast variables is assumed to be strictly uniform (this property holds for all pure energy eigenstates; it does \emph{not} hold for superpositions of these eigenstates). As soon as we add ripples in this distribution, this is tantamount to admixing higher energy modes. We use this observation in our discussion of interference, see  Appendix \ref{interfer.app}.

This will affect all other statistical distributions. We can use either quantum mechanical or classical formalisms to deduce what is the most likely thing to happen; the outcome of our calculations should not depend on how we perform them.

Important observations may be added concerning the question of quantum interference. In view of the importance of this issue we took it out of the main text of this paper by adding a special appendix to explain the situation, Appendix \ref{interfer.app}.

Thus we conclude that all that is needed to turn a classical system into a quantum mechanical one, is to define energy as it is only done in quantum mechanics, that is, by diagonalising the evolution operator. After this, one has to postulate that the fastest moving parts of the system must be limited to their lowest energy states. Effectively, this amounts to slightly smearing the amplitudes in the time direction in order to account for the limitations in our time resolution.

Besides possible modifications of the Standard Model in its highest energy domains, there may be implications for investigations of cosmology. When the universe was very small, the energy density must have been very high. Maybe all energy states were equally occupied when the universe started as a single point, or almost as a point. At the indivisible instant of the Big Bang, there was no quantum mechanics yet. The universe expanded, and this forced it to cool off. Subsequently, the laws of thermodynamics deprived our world of its highest energy states, with quantum mechanics as a result.

We also see implications for quantum black hole physics. When a black hole forms, imploding matter gets compressed against the past event horizon, as seen by a  distant observer at later times. Similarly, outgoing Hawking particles line up along the future event horizon, ready to spring to life much later. When matter reaches very high energy densities this way, it may well be that the \emph{highest energy state possible} is approached near both horizons. This state contrasts with the state with lowest possible energy density, the vacuum state.  We called it the `antivacuum' state. In the classical picture, a symmetry relating vacuum to antivacuum seems to be evident. When we describe stationary black holes, matter appears to be almost absent, as if the antivacuum of compressed imploding particles has been transformed into a vacuum. Since matter is the source of curvature, this replacement of antivacuum with vacuum forces the past and future horizons to change their effects on space and time. This is where the `antipodal identification' is suspected to originate. Antipodal identification is known to be needed if one wants to restore unitarity for the evolution of a stationary black hole.\cite{GtHBHantipodal.ref} 

Note that, in crossing the event horizons, one lands in domains where the sign of the Hamiltonian is inverted. Indeed, this corresponds to a transition where vacuum and antivacuum are interchanged, and it also answers an objection raised by E.~Witten to the author: the \(i\) in Schr\"odinger's equation turns into \(-i\) when you invert time; should this not invalidate the antipodal identification? Our answer is no, if we take care of also inverting the particle - antiparticle population while crossing a horizon.\fn{E. Witten, private discussion.}

Note also that, for the fast classical variables, there is hardly any distinction between the lowest energy states and the highest energy states.
 
Other questions are also still wide open: for instance, we wish to explain the existence of quite a lot of continuous, global and local symmetries of our world. Making discrete, classical theories that respect these symmetries (gauge symmetries, Goldstone symmetries, special and general relativity, supersymmetry perhaps, and so on), is notoriously difficult. 

We do mention in passing that there may exist interesting \emph{approximate} symmetry transformations in nature's cellular automaton. These symmetry operations will not quite commute with the Hamiltonian, but instead, act much in the way of, bosonic or fermionic,  physical particle fields. One may wish to conjecture that this is what the particle fields of the Standard Model really are; they describe the known particles as low energy excitations. Through the Goldstone theorem, the masses of the physical particles are indicative for the violation of these approximate symmetries. And as these masses are indeed very low as regarded in the Planck regime,  we are dealing with very interesting approximate symmetries.

Such and other questions are left for future investigations.

 In our model, all slow observables \(|i\ket\,,\ |j\ket\,, \ \dots\) are ontological, but the fast ones are put in an energy eigenstate, which is ontological, or more precisely, the total energy of the entire universe is declared to be ontological.\fn{We say this because the total energy is conserved, but we have to realise that total energy does not commute with the other ontological observables, and normally this would not be allowed. Here, we see no problems in constraining the total energy, but a more precise justification would be welcome. For instance, rather than imposing any absolute constraint on total energy, one may demand energy to lie within some domain. By respecting the energy vs time uncertainty relation, smeared energies and time-smeared observables may all be called ontological.} The computational rules are as in ordinary quantum mechanics, but the entire theory is fundamentally deterministic. Statistics enters at the moment we single out the ground state for the fast variables.

At first sight, our model differs widely from the Standard Model, but it does seem to be built from variables that may be regarded as quantum fields. One suspects then that these quantum fields are separated into slow and rapidly moving components, which may well take discrete values when regarded at the Planck scale. We see that refined restructuring of nature's degrees of freedom at the ultimate physical length and time scale, will be inevitable; model building with such ideas in mind will be left for future, more advanced investigations.

Regarding our observation that models of the sort described here are perfectly guaranteed to represent pure quantum mechanics, it may well be that a strict separation between fast and slow modes is unnecessary. Even the Standard Model admits, and indeed favours, the existence of ultra heavy particles. The vacuum fluctuations of their quantised fields are perfectly suitable to play the role of fast variables, and this is why we suspect that, indeed, quantum mechanics generated in line with our description, is almost inevitable. Thus, our theory also explains \emph{why} we have quantum mechanics.

In Appendix \ref{entsym.app}, we speculate further on the emergence of symmetry groups and low mass particles in the Standard Model.

We conclude that quantum mechanics may well be perfectly understandable if the right mathematical framework is used. For this author, the present results turn quantum mechanics as it was known in the literature, into a completely conventional, yet complex, classical system. The fastest variables take values that can best be described as `white noise'. The slower variables react by taking values that can only be understood as statistical data. What is presently known as `quantum mechanics' is nothing more than a machinery for the optimal treatment of the statistical rules.

An intriguing observation is furthermore that the quantum field variables appear to be constrained to lattices (the locations \(x_{i}\) of the  fast variables in our model). This forces also the interaction constants of the resulting theories to lie on lattices; they are not continuously adjustable. This is why one may suspect that investigations of this type may be useful for the future of model building.\\[10pt]

{\noindent {\Large\textbf{\noindent Acknowledgment }}\\[10pt]
The author benefited a lot from discussions with T.\,Palmer, F,\,Scardigli, C.\,Wetterich, A.~Schwarz, and many others. \\[10pt]

\appendix

\noindent{\Large\textbf{Appendices}}\\
\newsecl {Quantum interference}{interfer.app} 

Since we derived quantum mechanics, it is obvious that quantum interference must occur, but how can it be explained that quantum solutions for wave equations can be superimposed in a way that (constructive or destructive) interference takes place, even if single events are considered? We claim that interference is a direct consequence of limiting ourselves to the lowest energy state  of the fast variables. This is tantamount to postulating that if we start with a probability distribution for the fast variables that is statistically strictly even, we shall see interference patterns arising. The zero energy state has maximal overlap with the completely even distribution, which was essential for the emergence of terms in the Hamiltonian that produce superimposed states. Thus we predict that an interference experiment can be mimicked when this model is tested in a computer simulation. 

But then, an objection can be put forward. Consider any standard interference experiment with light or particles going through two slits. What puzzles people most is that closing one slit may actually enhance the probability of a particle hitting a screen at some place downstream. What happens inside my classical computer? In the computer, one could have registered which slit the particles pass through, without disturbing the outcome of the other calculations. Suppose we perform an ideal interference experiment. After having admired the beautiful interference pattern, we select out all cases where the particle was seen to pass through one slit, and compare that with the cases where the particle went through the other slit. 
Both of these selections should not show interference, because the particles went through one slit only. This obviously cannot happen in a classical computer, since recombining the two sets should give dark spots where no particles arrived at all.   What did we do wrong?

Indeed, the argument that, since the particles went through a single slit, there now should be no interference, in this case is wrong.

What really happens can de derived from quantum mechanics. Since none of the particles could have arrived at the dark spots, selecting out all particles that went through one given slit indeed will not remove the interference pattern.  It does something else however: if we select one slit, and repeat this experiment many times, then we are making a selection among the initial states chosen for the fast variables. This selection will not be an even one! Therefore, the initial state of the fast variables must now be described as a superposition of different energy eigenstates. Since now the particles went through a given slit, each energy eigenstate of the fast particles should not give any interference pattern. However, we now are describing an experiment where the initial state was a superposition of at least two energy eigenstates of the fast variables! Superposition \emph{in} = superposition \emph{out}. Thus, the observed interference pattern is due to interference between two differently chosen initial states of the fast variables, and not due to the slits. This explains the result.\fn{Of course, one can continue asking questions: \emph{Why then, does the interference pattern vanish when a physical observer  checks the particles at the slits?} There again, ordinary quantum mechanics gives the answers. A physical observer cannot affect the fast variables, he does not have the energy to modify their statistical distribution, while for the computer this had been no impediment at all. To close one of the slits, the physical observer has to modify the Hamiltonian by changing the openings; (s)he cannot do this by putting the equivalent of Maxwell's devil near the slits.}

In short, the answer comes from the small-print: limiting ourselves to the lowest energy eigenstate of the fast variables was imperative. The fast variables have to start in a perfectly even distribution. Selecting out particles that took specially chosen orbits disturbs the even distribution of the fast variables.

\newsecl{What happens in an EPR/Bell experiment?}{EPRBell.sec}
	The critical reader may have thought that there is one thing that will fail to be clarified: \emph{What happens in Bell's Gedanken experiment} when Alice and/or Bob choose their settings to record the spins of the photons they observe, and how can this be reconciled at all with locality and causality? Bell, and most of his followers up to this day, were convinced that the quantum mechanical prediction of the observations, will clash with all classical descriptions. Yet here we have a classical theory claiming to mimic, as precisely as one may desire, what will be observed. We shall not repeat the numerous discussions in the literature, but just tell what happens, and where the `totally natural assumptions'  made to arrive at Bell's theorem, fail badly: 
	\begin{quote} The way Alice and Bob decide to choose their settings depends in the state of the atom when, in its far past, it emitted two photons.\end{quote}
This was called `superdeterminism', or `conspiracy', or the `failure of free choice'. All these were thought to be unacceptable. But this `clash with common sense' has a natural explanation. It comes from disregarding the small print, as it was mentioned already in Ref.~\cite{GtHCA.ref}, Chapter 4.2 (deterministic Quantum Mechanics). There, it is emphasised that, and explained why
\begin{quote} all ontological  states in the macroscopic world, are also ontological in the microscopic sense, and vice versa. \end{quote}

Now this surprised people. \emph{Why, oh why?}, exclaimed one referee. To the present author, this question just shows how strongly we have been indoctrinated by the Copenhagen way of thinking during almost one century. The macroscopic classical limit was generally thought to arise through some form of `decoherence', which should be a large scale feature, having little to do with the microscopic world. 

But if we just focus on the classical description of the evolution law, in our terminology, it is clear what happens instead. Our classical law does not know about quantum mechanics, let alone superposition, Schr\"odinger's cat would just be a classical cat, and so on. 

Look at Bell's description of his experiment\,\cite{Bell1982.ref}. How do we formulate what is going on in our classical terms? To describe this appropriately, we need an orthonormal basis of states (labelled by \(i=1,\,\dots, N\)). We wish our choice to lead to local equations. The best basis is then the  set of values of the vector potential\,\fn{There is a subtlety with photons: the vector field \(A_\m(\vec x,t)\)  is not gauge-invariant. As long as we stick to one given Lorentz frame this is no problem, and in the case of photons all one has to do is replace it by the magnetic field. Yet a general discussion of gauge invariance in this theory
will be needed.} field \(\vec A(\vec x,t)\), whose values at given time all commute. In this basis, our procedure generates the correct Schr\"odinger equation.  The probabilities thus calculated should precisely match the physical situation.

    The vector potential does not commute with its time derivative, which implies that the time derivative depends on fast variables -- but, together with its fast variable, the vector potential is still ontological.

As long as Alice and Bob keep their settings as they may have chosen them, there is no contradiction. But now suppose either Alice or Bob change their minds at the last moment, without admitting any change in the past. In quantum mechanical terms, the photons they now are looking at, would be quantum superpositions of the photons they had before. Now  \emph{that} makes no sense in the classical description. It would mean that, extrapolating back to the initial state, we would be looking as some superposition.

It is not difficult to accept the notion that, even minor changes in Alice's or Bob's settings, would require some modifications in the classical past. That would be necessary in any deterministic theory. Now, we see that \emph{there is no way we can modify the classical initial state such that the observed photons in the future will be in the required superposition.} Instead, some classical change has to be made in order for an observed photons to end up in either a state where its observation yields a \emph{yes},  or it should yield a \emph{no}. Superpositions are not an option. 

The fast variables are a classical probabilistic mix, so that our initial state allows for many possibilities for the outcome. Thus, the whole experiment is predicted correctly by the mimicked quantum mechanical expressions. This includes both the settings chosen by Alice and Bob, and the classical photons that will be observed. However, if we just stick to the classical formulation, all probabilistic distributions that ever arise -- both the photon distributions and the settings chosen by Alice and Bob -- are due to the strictly even probabilistic distributions of the fast variables in the initial state. In the classical picture, at this point, there is no question of conspiracy anywhere.

\newsecl{Entanglement and symmetry}{entsym.app}
When we try to imagine a cellular automaton that would mimic the Standard Model of the subatomic particles any way, then we encounter a real mystery: its Lie symmetries.
As was stated in the introduction, it is hard to imagine where and how such symmetry structures can originate in a cellular automaton. A symmetry is to be defined in two steps: \\ One, describe a transformation among the (discrete) data occurring in each cell, such that the transformed system evolves almost in the same way as the original. This is easy if one considers translations in space and in time, or rotations under \(60^\circ\) or 
\(90^\circ\), but already rotations over smaller angles will be harder to understand. We have transformations affecting the identity of particles (isospin), but also transformations replacing particles in a space-time dependent way (local gauge symmetries), and finally transformations that replace space-time coordinates by curved coordinates (which are known to be relevant for understanding gravitation).\\
Two, compare accurately the evolution laws of the cellular automaton before and after the transformation. One might find that the effective Hamiltonian stays not quite the same, but gets an interaction term added to it that can be interpreted as the mass term of a field equation. The field corresponds to the (space-time dependent) transformation. \\
But we can now  make a third step:  note that, if we choose the space-time dimensions as being of the order of the Planck scale, we see that many of the mass terms in the Standard Model are very small. This means that, in Planck units, the changes brought about by the transformation field may be very tiny, and one might deduce that, at the Planck scale, every field of the Standard Model represents a space-time dependent symmetry generator. Bosonic fields in the Standard Model are generated by real numbers in the generators of a symmetry, while fermions may be associated to binary transformations, that is, transformations that square into one, so that they may resemble supersymmetries; thus we imagine that each and every quark and lepton species is to be associated to some distinct supersymmetry generator.

We mention entanglement in the title of this appendix just to suggest that the generators of these local transformations may easily get entangled into one another.
This subject has not yet been studied further, but we suggest that it could lead to indications as to what shape the Standard Model will take at Planckian scales.

\newsecl{On the quantum computer}{qucomp.app}
Our theory suggests a straightforward construction of a classical model for mimicking `quantum computation'. Quantum computations become of interest since it is suggested\,\cite{Shor.ref} that they should be able to solve non-polynomial problems fast and efficiently, basically in polynomial time. At first sight this appears to lead to contradictions, as our classical system seems to be able to lead to the same solutions as well, while it is polynomial. 

We did indicate that, for every single state \(|i\,\ket\) of the quantum system that we simulate, a classical variable is required that can be in a large number of states, but this does not quite remove the apparent contradiction.

There are a few obvious observations to make:
\bi{1)} We have fast and slow variables. When running a simulation program, the fast variables will not be able to run faster than our classical computer can handle, which will imply that our slow variables will have to run slowly indeed, and it is the slow variables that will do the real `quantum' calculation. To realise a simulation process, it will have to be made to speed up, in order to do the purported `quantum computation' really fast.
\itm{2)} Our operators \(\s_a\) are ideally suited to serve as qubits in a pseudo-quantum calculation.
\itm{3)} Careful examination of Shor's algorithm, does raise concerns about its applicability for either factorisation or the determination of 
discrete logarithms of large numbers. It is claimed that a quantum computation can speed up the determination of the exact period of a sequence of almost random numbers. Classical approaches will fail for the simple reason that, regardless the algorithm used, a very large subset, roughly \(\sqrt{N}\) of these numbers, have to be compared before two equal numbers are encountered, so that their separation betrays the period. If the sequence had been smooth then periodicity could be established more efficiently, but, at least to this author, it is not clear whether this will work for large, almost arbitrary, sequences of integers.
\itm{4)} If one tries to speed up our simulation, for instance by choosing smaller values for the periods \(L_i\), one might have to deal with higher order effects that are difficult to calculate; it may well be that these will end up acting as agents of decoherence, as seen by the simulated observer. This is unclear to the author at present. Needless to say, however, that this will be one of many alleys for further investigation.\ei

 \end{document}